\documentclass[prb,preprint,floats]{revtex4}
\usepackage{graphicx}

\begin{document}


\title{Developments in Correlated Fermions}
\author{C. M. Varma}
\address{University of California, \\
Riverside, CA 92507}

\begin{abstract}
This manuscript is based on the  Summary and Overview talk given at the "The International Conference of Strongly Correlated Electronic Systems" (SCES '04), July 26-30, at Karlsruhe, Germany. After highlighting some of the principal new experimental developments in heavy fermions presented at the conference, I turn to two kinds of theoretical questions. (1) What is understood of the fermi-liquid state of the heavy fermions and what is not, but may reasonably well be understood by systematic calculations. (2) The profound issues raised by the observed correlations near the quantum critical points in the heavy fermions. The numbers and letters in the parenthesis in the text refer to the listing of the talks in the "Program and Abstracts" book of the conference.
\end{abstract}
\maketitle
\subsection{Introduction}

I am very grateful to Hilbert von Lohneysen and Peter Wolfle for giving me the opportunity to speak of the progress and the important questions in Correlated Fermions, especially the heavy Fermions, and to put in a broader framework some  of the many important papers dealing with specific topics presented at this conference. I was a very active early participant in the development of this subject in the 70's and 80's. With the advent of the Cuprate problem I have been watching this subject from the sidelines, although this field and that of the cuprates  have had important influence on each other.  Frank Steglich reminded me that I had summarized the developments on a conference on heavy fermions in Cologne twenty years ago, in 1984. So, it has been a very rewarding experience to attend most of the talks at this meeting and visit the posters and think of the developments in this field and the problems remaining.

What have been the truly remarkable developments in this field since about 1990? Leaving aside the Cuprates, in my opinion, they are the discovery of the Quantum critical points in the heavy Fermions and the realization that this is truly unchartered territory, the discovery that there is a metallic state in two dimensions, the recognition that there are "hidden" order parameters in heavy fermions (and perhaps other metals) which are not simply the easily discoverable change of translational symmetry or spin-rotational symmetry or superconductivity, and the growth of the field of frustrated magnetism both of the classical and the quantum type.

What of the developments in the theory of these phenomena. In my opinion this has lagged behind. The principal new developments are the recognition that the QCP do not fit into the nice easy extensions of classical critical phenomena done in the 70's and 80's and put in the language of renormalization group by  Hertz, Beal-Monod and Maki, and by others. A new idea moving the field, is that the singularities at the QCP may be due to local physics. I shall comment on this idea and its implementations below. There is also the rapid growth of non-perturbative field theory methods to solve model problems and to establish general principles which may have applicability to the models relevant to experiments. These were exemplified at this meeting in the lucid talk by Subir Sachdev (MO-P II and elsewhere). The principal difficulty here is that in our field, it is hard to solve any significant model which includes Fermions, by such methods. 

The other great progress has of-course been the development of the Dynamical Mean-field method  which with judicious handling (Kotliar MO-P III) has become a growth industry. 
With the extensions to cluster DMFT this method is the most powerful way to address questions in solids since the advent of the methods of band-structure calculations on the one hand and the RPA methods on the other. With modern computers it is probably no more time-consuming than good bandstructure calculations were 25 years ago. As Kotliar mentioned, problems involving criticality and large length scales are not  answered  by this method. But even for such problems, cluster DMFT holds the promise of deciding the relevance of different interactions in the  physics beyond mean-field theory. 

The plan of my talk is as follows:  I will first  highlight  some of the promising new experimental developments presented at this conference which I think will influence work in the next decade. Then, in order to be able to say something besides polite generalities, I will confine my remarks to the developments reported at this meeting on heavy-fermions, including QCP's.  I will speak of
the essentials of the heavy-Fermions which we do understand and which we do not understand but which I think are likely to be understood. I think this is necessary before we go to the next step: discuss problems in which we have at the moment only a faint glimmer of understanding. These concern QCPÕs, and I would like to raise some questions whose answers I would like to know. I would also like to put these remarks in the historical context that these questions arose.

\subsection{Some interesting experimental developments}

With this narrow scope, the developments in experiments which particularly impressed me and which in my opinion will influence our thinking and our work for at least the next decade are
the beginnings of deHaas van Alphen and ARPES methods to probe on both sides of the transitions at QCP's and elsewhere. This was in evidence in the talks of Julian (MO-P I) and of Allen (TU-P IV) at the meeting. Also, the investigation of Hidden order parameters by ARPES, alluded to by Allen and  by
Denlinger (TH-ACT2-44). These microscopic probes, when fully developed for such problems  have a lot to offer: the study of the changing one-particle self-energy as the QCP is approached and the nature of the gap in the hidden order parameter case; by deHaas van Alphen, the changing shape and size of the Fermi-surface. Is it also possible to develop the technique of spin-polarized ARPES?

The discovery of superconductivity in heavy-Fermion metals which break the center of symmetry (Bauer et al.,TU-RES1-1, and several other contributions at the conference) should spawn a lot of interesting group theoretical work (Frigeri et al. TU-RES1-3) . In this connection, the fact that these materials have fully developed AFM is not something which can be ignored. Quite generally an s-wave superconductor in an AFM must have a triplet order parameter mixed in. Also, interesting is  a material with three superconducting transitions, ( Cichorek et al. FR-RES8-88) whereas two have been known for quite some time. It is actually a cubic crystal, so the phenomena has obviously to do with the higher dimensional irreducible representations possible in a cubic crystal.
The discovery of superconductivity in Fe at high pressures (Jaccard and Holmes WE-RES5-2) is most amusing and given the phase diagram and the evidence for ferromagnetic fluctuations
has at least as much grounds for being considered a triplet superconductor than any other compound I have come across. there is some cleaning up to do, which may be hard since a first order transition from ferromagnetic to paramagnetic phases is close by in pressure. 

The discovery that $CeCu_2Si_{2-x}Ge_x$ has two regions of superconductivity, (Steglich WE-P5, Jaccard WE-RES5-2), one associated with an AFM critical point and the other in a region of valence change is noteworthy because it tells that d-wave supreconductivity can arise from more than one kind of fluctuations as emphasized by Miyake and others.

The discovery (TU-QCP-20) of a QCP in a cubic material, $CeIn_{3-x}Sn_x$  with all the usual signatures $ ( \rho \propto T, C_v/T \propto \log T)$ puts a constraint on theories which regard such properties as having to do with two-dimensionality of the spin-fluctuations. It should have an important effect regarding the generalities of the observed anomalous properties, which at first glance seem similar to those in $CeCu_{6-x}Au_x$.

I was also struck with the discovery (Ulharz TU-TMC-38) that when  the ferromagnetic transition temperature in $ZrZn_2$ is reduced by applying pressure, it changes to a first order transition just as it does in $MnSi_2$. Are all ferromagnetic transitions at low enough temperature first order?

 In the experiments of the Dresden group (TU-QCP-13) on $YbRh_2Si_2$, the ordered moment at low fields is less than 0.04$\mu_B$, and the magnetic order at zero-filed sets in at a temperature of $ 70mK$. But the fluctuation regime where anomalous properties associated with the QCP are observed is of order $10K$ and the magnitude of the fluctuating moment $<M^2>$ is of order $ 1 \mu_B^2$. The magnetic order appears incidental except to define the QCP. Under pressure the zero-field AFM transition indeed occurs at a few kelvin and there is a line of critical points in the P-H plane. But how does the ambient pressure compound sample the amplitude of fluctuations characteristic of those at much higher pressures?

\subsection{What do we understand about Heavy-Fermions and what we don't}

To put in context my remarks, let us look at what seemed to be an unqualified success in our understanding of heavy-Fermions, the hybridization diagram of fig. (1). Basov (TU-TM01-01) talked about how well the optical spectra (long-wavelengths) fits the predictions of this picture and McEwen (WE-ACT1-4) of how neutron scattering gives results over the whole range of wave-vectors giving agreement with this picture for both charge and spin-fluctuations. What is the basis for this picture and how much understanding of heavy Fermions does this represent? The physics in fig.(1) and the figure itself first appeared  in 1976 \cite{cmv-yafet} where the hamiltonian, now popularly called the Anderson lattice Hamiltonian, was first introduced. Subsequently, this has been derived also by the $1/N$-methods, slave boson methods \cite{slavebosons} and Gutzwiller \cite{gutzwiller} approximation. It represents the most rudimentary understanding of heavy Fermions leaving many subtle questions unanswered. Let me first talk of the context in which this was derived.

       \begin{figure}
\includegraphics[angle=0,scale=0.5]{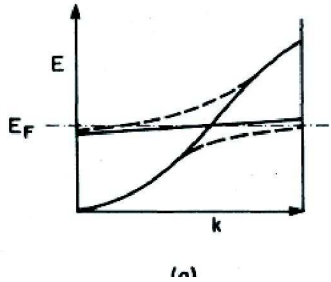}
    \caption{The "band-structure" of mixed-valence and heavy Fermions, taken from
    C.M.Varma, Rev. Mod. Phys. {\bf 48}, 219 (1976). }
    \label{fig:figure1} 
    \end{figure}

The field of heavy-fermions started in the late 1960's and the 1970's through the work of Maple and Wohleben who began to study the magnetic properties of intermetallic alloys with Ce and Ted Geballe who investigated $SmB_6$ and Jayaraman and others who discovered metal-insulator transitions under pressure in the Sm-chalcogenides. The first surprise was that the magnetic susceptibility $\chi$ of these compounds in the metallic phase was of the Pauli type unlike the majority of metals with rare-earths and second that it was high matching the measured $C_v/T$, giving a mass enhancement of about 50-100. It was clear that the usual f- local moments had been replaced by itinerant f-electrons. It was also known that in Ce and Sm and Yb, the energy distance between the last occupied 4f level and the Fermi-energy in a band of 5d and 6s elecrons is particularly small providing the prerequisites of a large exchange-energy between the two.

 It was about this time that the Kondo effect had just been understood through the work of Anderson and Wilson and others, after a decade long effort. Nozieres had just synthesized the solution in the language of Landau Fermi-liquid theory and we had understood that well below the Kondo-temperature, the local moment acquires the itinerant character of presenting a temperature independent $\chi$ and $C_v/T$. But first we had to understand the relationship of the Kondo temperature to hybridization through which one could construct band-structures. Yafet and I first did a variational calculation on the Anderson impurity  model and showed that the ususal expression for the Kondo temperature could alternately be expressed as
\begin{eqnarray}
T_K= \gamma(1-<n_{f,\sigma}>).
\end{eqnarray}
\label{tk}
Here $\gamma$ is the usual one-electron hybridization of the impurity level to the conduction electrons and $(1-<n_{f,-\sigma}>)$ expresses how much the hartree-fock resonance is unoccupied. (Kondo effect is just constrained hybridization; a conduction electron of spin $\sigma$ can hybridize with a local level with a very large repulsion
$U$ only when the level is unoccupied with the opposite spin. $(1-<n_{f,-\sigma}>)$ expresses the probability of that occuring.) Next we considered the lattice model, appropriately decoupled the Hubbard constraint operators that arise and obtained the picture of fig. (1) where the hybridization parameter is this renormalized hybridization of Eq. (1) or $T_K$. No essential new physical result is found in the other methods mentioned above that followed till the DMFT which gives additional frequency dependent information and structure  through the self-energy.

Actually, fig. (1) has only one new piece of information, Eq. (1). The rest is simply the application of the Bloch-Wilson band structure rules for the number of bands of a solid in relation to  the number of orbitals per unit cell and the conditions to get a metal or an insulator in terms of the number of electrons per unit cell. What about the rest of the interesting physics which is present already in the Kondo problem, for example the Landau parameters and the additional scales which must arise when we have a periodic array of interacting Kondo atoms? The methods by which fig. (1) was developed are ill equipped to deal with these. So is the single-site based DMFT. I think a low energy Hamitonian for the lattice problem derivable from, for example, a cluster DMFT is required to answer these questions. I also think the considerations of critical points should be based on such a Hamiltonian; it is the instability of the excitation spectrum of such a Hamiltonian which is the genesis of the critical points.

In the compounds initially investigated $(1-<n_{f\sigma}>)$ was a few tenths. But in 1975  there came the truly Heavy Fermions; the first was $CeAl_3: (1-<n_{f\sigma}>) \approx 0.01$. This was \cite{andres} just after fig. (1) was derived so the heavy mass $O(10^3)$ was interpreted by these ideas as a Fermi-liquid whose physical basis and the heavy mass is the Kondo effect. 

What of the physics not contained in fig. (1)?
What are some of the specific questions which are unanswered for the Fermi-liquid state of the heavy Fermions and whose answers can be found with existing tools?:

(1) The length scale in the Kondo problem and the question of Nozieres exhaustion principle \cite{nozieres}: Arguments can be given that the Kondo scale $T_K$ corresponds to the length scale
$\xi_K=v_F/T_K$. This leads to the Nozieres irritating exhaustion principle, irritating because if it is correct, its resolution leads to energy scales in the lattice which are vastly different from that of the single impurity and which seem to me implausible compared with the empirical results. Irritating also because no convincing fault in the Nozieres idea has been found. True, arguments can also be given that the Kondo problem is all in the frequency domain and is a local problem spatially. My guess is that which of these point of views is correct depends on the question asked, more precisely what is the frequency scale at which one is asking for the correlation length? This should be answerable by an appropriate Wilson-RG for the correlation functions. But to my knowledge this has not been done.

(2) The question of the coherence scale: An idea of the basic physics of the coherence scale in the lattice should be obtainable from the complete solution of the two Kondo impurity problem. When Barbara Jones and I \cite{bajones} formulated this problem for solution with the Wilson RG, we investigated only the particle-hole symmetric case. This led to the discovery of a quantum critical point, which is unstable to particle-hole asymmetry. Not well known is the fact that it also
fixed the even and odd parity resonances of the two impurities in the Kondo regime tied to the chemical potential. With particle-hole symmetry asymmetry, which is the generic case, the two resonances would be split. See fig. (2). This splitting is the physics of the band-width in the lattice problem. The connection from the bare terms which give particle-hole assymetry to the splitting of the resonances in the asymptotic low energy hamiltonian has not been investigated. We know empirically that this should come out comparable to the Kondo temperature; it is important to know why?

      \begin{figure}
\includegraphics[angle=0,scale=0.5]{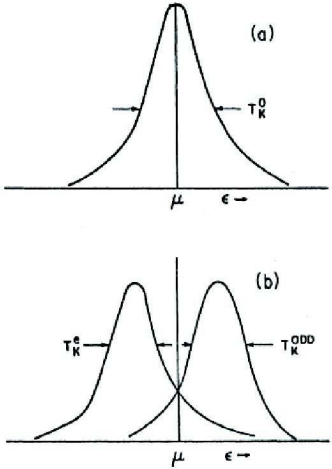}
    \caption{(a) The single-impurity resonance, (b) The correct answer expected for the two impurity resonances in the odd and even parity channels, presaging the "coherence" scale for the heavy fermion problem }
    \label{fig:figure2} 
    \end{figure}

(3) The Landau parameters for the problem: The Wilson solution of the two impurity problem can be couched in terms of a low energy Hamiltonian which gives the full set of interaction among the quasi-particles through the leading {\it irrelevant} operators. Included in these are the operators which are the analogs of the RKKY interaction as well as the pairing interactions. 

The question of the RKKY interactions and the competition with the Kondo effect ought to be discussed a bit. The principal surprise when the mixed-valence and heavy-fermion fields were discovered was that they did not order magnetically. The question was in the air, why? The answer was stated vaguely by me in my review article\cite{cmv-yafet}  that perhaps because of the Kondo effect at a temperature $T_K$, the fermions turned into a spin-liquid at a temperature higher than where  RKKY could be effective in turning them into a spin-solid. There was hesitancy in saying more because the energy scales are all wrong. Doniach was bolder \cite{doniach}. He drew a picture where as a function of $J\rho$, where $ J$ is the exchange interaction of the local moments with the conduction electrons and $\rho$ is the density of states of the conduction electrons, there is a crossing of the Kondo temperature $T_K$ and the RKKY interaction $I$,
\begin{eqnarray} 
T_K \approx \rho^{-1} \exp(-1/NJ\rho),~~~I \approx NJ(J\rho).
\end{eqnarray} 
The  hesitancy is that with N=1 for the $S=1/2$ Kondo problem, the crossing occurs at $J\rho \approx 1$, which is quite an absurd requirement. Subsequently Ramarkrishna \cite{ramakrishna} pointed out that N is the orbital degeneracy of the relevant f-level. The situation might appear to have gotten much better. But if one puts in the actual degeneracy in the lowest multiplet in the compounds studied, it is generally not high enough to understand why they are ever heavy Fermions and why they do not simply magnetically order. 

A possible answer to this question is the mundane fact that if the crystal-field multiplets are reasonably well separated, there is a multiplcity of Kondo temperatures with crossovers between them. This is clearly discerned when the heavy-Fermions are studied with the density of the rare-earth ions systematically varied, as done for example in one series of the so called $115-$ compounds \cite{nakafutsuji}.
There is a lot to be learnt from such dilution studies and I would strongly urge more of them. Incidentally, at temperatures in between the two Kondo scales, the lower multiplets would present the entropy and magnetization close to that of local moments. I suspect this is the reason why in the compounds like $YbRh_2Si_2$ studied by the Dresden group, the data is fitted with a fraction of the f-electrons localized and the rest itinerant. One should also bear in mind that the crossovers in the compounds between the Kondo scales are likely to be slower than in the isolated ions and if one is not careful, one might ascribe more fancy things to them.

Another (and  in parallel  to the above) answer is that the low energy Landau parameters providing the magnetic coupling are quite different from the bare parameters. One of the strange facts is that in the heavy Fermion regime the single impurity scale seems to govern almost all the physics and the other Landau parameters are only of order unity.
These things were realized very early through an attempt to understand the renormalization of the equilibrium and transport properties in heavy Fermions \cite{cmv}. The dominance of  the single impurity Kondo scale for the properties of the heavy Fermions is equivalent to the statement that the self-energy is independent of momentum. This is of-course the intellectual background to the growth of the DMFT method.

 If we wish to learn more as an answer to some of the questions raised above and especially if we wish to investigate the instabilities of the heavy Fermion state, one must derive a low-energy hamiltonian for the lattice from one of the generalizations of the DMFT methods to treat a cluster of f-ions self-consistently. Such a  Hamiltonian will contain information on the coherence scale, the Landau parameters and the leading operators which lead to magnetic or superconducting instabilities of the heavy-Fermion state. I think it is also a first step towards thinking on questions about quantum criticality.

\subsection{Theory of Quantum Criticality}

I will frame my remarks on the theory of QCP's  in terms of some questions which have arisen in my mind  
based on several of the theoretical and experimental talks at this conference:

 (1) How many classes of quantum critical phenomena exist? Is there any material for which the RPA theories of Moriya \cite{moriya} and Hertz \cite{hertz} etc. work? A part of this question is, does $\omega/T$ scaling work in every experiment. I have been receiving diverse opinion on this. We know three examples of observed $\omega/T$ scaling. In cuprates \cite{slakey} through Raman scattering experiments, i.e. for $q\rightarrow 0$; the results in neutron scattering \cite{stockert} near the AFM wave-vector in $CeCu_{6-x} Au_x$ and the neutron scattering results \cite{aronson} over a range of wave-vectors in $UCu_{5-x}Pd_x$. There also exist some experimental results which suggest that at some critical points $\omega/T$ scaling appears not to hold \cite{kambe}.
 
 $\omega/T$ scaling was first proposed \cite{mfl} for the cuprates, as the form of fluctuations which must occur to account for the diverse anomalous normal state (marginal fermi-liquid)  properties in the cuprates near the doping concentration for the highest $T_c$. Sachdev \cite{sachdev}  pointed out that this form cannot arise in a Ginzburg-Landau type  theory with bosonic variables alone (such as the theory of Moriya and Hertz) if  $d+z$ is larger or equal to $4$, when the $\psi^4$ interactions of the slow variables in terms of which the theory is framed become irrelevant. Since $d+z >4$ in all the systems in which $\omega/T$ scaling is observed, something is seriously amiss with the foundations of such a theory. This is especially serious because the theory is consistent; given its starting assumptions, which are straight-forward generalizations of the successful theory for classical critical phenomena, the corrections to the theory are unimportant.
 
 (2) To get around this problem, the starting assumptions must be challenged. One idea which implictly does so is the idea of local quantum criticality. 
 
  The concepts behind local quantum criticality were first introduced in connection with the cuprates \cite{cmv1}, though the name was invented by Qimiao Si \cite{si} in connection with $CeCu_{6-x} Au_x$ and a specific calculation provided to obtain a set of appealing results. By local criticality is meant  that the cause of the singularity can be found in a local problem and that this singularity does not change in an essential fashion in a lattice where there is an interaction between the periodic array of such local problems. Such interactions are "irrelevant". Does this idea of local criticality have validity and if so under what conditions?
For most of the impurity critical points we know, for example the multichannel Kondo problem, two interacting Kondo problem etc., the interactions change the special symmetries required for the criticality. So, if local quantum criticality exists, there should be some conservations laws or other general principles, which protect the criticality. I dont think such general principles are known yet? 

A new idea (Sachdev MO-PII) , not completely orthogonal to the idea of local quantum criticality is that the Kondo effect may be destroyed at a different temperature than the occurrence of magnetic order, except in the limit $T\rightarrow 0$.  In between the two
temperatures the magnetic moments turn to a spin-liquid distinct from a fermi-liquid. This state FL* has some rather special properties and so fortunately can be verified or ruled out, for  example by the deHaas van Alphen experiments.

(3) Can  Fermions be eliminated? The inapplicability of the Moriya, Hertz type  theory may be related to eliminating the Fermions in favor of slowly varying fluctuations, such as the magnetization for a ferromagnetic QCP or the staggered magnetization for an antiferromagnetic QCP, which may be treated as a bosonic fluctuations. But in a metal, this fluctuation is  a fermionic particle-hole fluctuation and its commutation rules are not  those of bosons. For many problems we may neglect the difference but it appears to me that this may make the crucial difference in this case. There is a specific problem I have looked at with Maebashi and Miyake: Kondo effect in a host near an antiferroamgnetic instability. 
The coupling of the afm fluctuations treated as bosons gives quite a different answer than treating them as objects with fermionic particle-hole legs. Vojta and collaborators at his meeting (TU-QCP-26) have come to similar conclusions in some other models.

(4) Is the Landau-Ginzburg-Wilson (LGW) type theory violated? This question is borrowed from the recent work of Senthil and collaborators \cite{senthil} where they found a model in which transitions to two different order parameters occur simultaneously which are not linearly coupled. What class of quantum-mechanical models would generally be outside the LGW framework is an open question. 

There are various different issues raised by this question. Let us look at the empirical side.  The scaling form with which the measured $\chi"(q,\omega)$ is fit, say in $CeCu_{6-x}Au_x$, obeys $\omega/T$ scaling which cannot be obtained from the LGW theory of Hertz.  This by itself does not allow us to say that LGW is violated because we do not have a LGW theory yet which keeps the slow fermions variables.  Certainly, the theory of local quantum criticality, right or wrong, is also a LGW type theory. 

 Could the Landau-Ginzburg- Wilson scheme break down in the sense that the coefficients of the $\psi^4,\psi^6$ and higher terms develop singularities as a function of the frequency or deviation from the ordering wave-vectors? Abanov and collaborators (private communication) have found such cases for AFM ordering in 2-$\epsilon$ dimensions. For commensurate problems, afm or cdw, at least the $\psi^4$ term also has singualrities as a function of $q-Q$.
 
 The other question is understanding the nature of the damping of the critical mode. My guess is that the nature of damping is pretty much fixed if the order parameter associated with the QCP is a conserved quantity. But if it is not, the situation is a lot more interesting. This problem is then related to how quantum-mechanical problems acquire dissipation in contact with a classical reservoir. I think the answer to this is not known except in some special cases.

(5) Can we classify Quantum transitions like Halperin and Hohenberg \cite{halperin} did for dynamics of classical phase transitions in terms of general properties of the order parameter. Are the universality classes of phase transitions different for problems involving slow fermionic variables? Do fermions introduce topological terms in the free energy making the LGW approach inapplicable. We know, for example, that the critical points of the impurity problems cannot be handled by LGW for this reason.

(6) Is there ever truly a $T=0$  critical point, or is it always obviated by an ordered phase or by a first order transition. The fluctuations near the QCP are soft enough that a transition to some ordered phase might intervene. But such ordered phases have their own putative QCP's. So one runs in to the crazy possibility of a sequence of nested phases as one goes lower and lower in temperature. 

In this connection an important question is what parameters set the scale which determines the observability of fluctautions for $T\simeq 0$ phase transitions? I have mentioned this question earlier in connection with the Dresden experiments where the effect of fluctuations are observable at temperatures far beyond what one would expect. The other side of the coin to this example is the question: why nothing peculiar seems to be observed in the quasi-classical regime of the  the superconducting quantum critical point.

{\it Acknowledgements:} Besides the very large number of people I had the fortune to talk to during the conference to whom I owe a debt of gratitude I must  mention conversations with Subir Sachdev, Qimiao Si, Frank Steglich, Mathias Vojta and Peter Wolfle as being especially useful to me in forming the views expressed here.

\end{document}